\documentclass{jfm}
\usepackage{graphicx}
\usepackage{float}
\usepackage{amssymb}
\usepackage{epstopdf, epsfig}
\usepackage{yfonts}
\usepackage[colorlinks=true, linkcolor=blue, citecolor=blue, urlcolor=blue]{hyperref}
\usepackage{fontenc}
\usepackage{upgreek}
\usepackage{mathrsfs}
\usepackage{booktabs}
\usepackage{pifont}
\usepackage{amsmath}
\usepackage{bm}
\usepackage{wasysym}
\usepackage{caption}
\usepackage{todonotes}
\usepackage{subcaption}
\usepackage{euscript}
\usepackage{natbib}
\usepackage{lipsum}
\usepackage{xcolor}

\usepackage{orcidlink}
\usepackage{amsfonts}

\usepackage[export]{adjustbox} 
\usepackage{cleveref}
\usepackage[normalem]{ulem}

\title{Lifetime Characterization of Extreme Wave Localizations in Crossing Seas}

\author{Yuchen He\aff{1,2}
  Jinghua Wang\aff{1,3,4} \corresp{\email{jinghua.wang@polyu.edu.hk}},
  Jingsong He\aff{5},
  Ye Li\aff{2},
  Xingya Feng\aff{2},
 \and Amin Chabchoub\aff{6,7}}

\affiliation{\aff{1}Department of Civil and Environmental Engineering, The Hong Kong Polytechnic University, Hong Kong SAR
\aff{2}Department of Ocean Science and Engineering, Southern University of Science and Technology, Shenzhen 518055, China
\aff{3}Research Institute for Sustainable Urban Development, The Hong Kong Polytechnic University, Hong Kong SAR
\aff{4}Shenzhen Research Institute, The Hong Kong Polytechnic University, Shenzhen 518057, China
\aff{5}Institute for Advanced Study, Shenzhen University, Shenzhen 518060, China
\aff{6}Disaster Prevention Research Institute, Kyoto University, Uji, Kyoto 611-0011, Japan
\aff{7}Department of Infrastructure Engineering, The University of Melbourne, Parkville, Victoria 3010, Australia
}

\begin{document}

\maketitle

\begin{abstract}
Rogue waves (RWs) can form on the ocean surface due to quasi-four wave resonant interaction or superposition principle. Both mechanisms have been acutely studied. The first of the two is known as the nonlinear focusing mechanism and leads to an increased probability of rogue waves when wave conditions are favourable, i.e., when unidirectionality and high narrowband energy of the wave field are satisfied. This work delves into the dynamics of extreme wave focusing in crossing seas, revealing a distinct type of nonlinear RWs, characterized by a decisive longevity compared to those generated by the dispersive focusing mechanism. In fact, through fully nonlinear hydrodynamic numerical simulations, we show that the interactions between two crossing unidirectional wave beams can trigger fully localized and robust development of RWs. These coherent structures, characterized by a typical spectral broadening then spreading in the form of dual bimodality and recurrent wave group focusing, not only defy the weakening expectation of quasi-four wave resonant interaction in directionally spread wave fields, but also differ from classical focusing mechanisms already mentioned. This has been determined following a rigorous lifespan-based statistical analysis of extreme wave events in our fully nonlinear simulations. Utilizing the coupled nonlinear Schrödinger framework, we also show that such intrinsic focusing dynamics can also be captured by weakly nonlinear wave evolution equations. This opens new research avenues for further explorations of these complex and intriguing wave phenomena in hydrodynamics as well as other nonlinear and dispersive multi-wave systems.
\end{abstract}

\section{Introduction}\label{section-intro}
Since the recording of the New Year or Draupner wave in 1995, fundamental research related to ocean rogue wave (RW) investigation has attracted much attention in recent decades due to its key relevance in coastal, ocean, and arctic engineering applications \citep{kharif2008rogue,osborne2010nonlinear,ducrozet2020experimental,mori2023science,toffoli2024observations,klahn2024heavy}.
Assuming the wave being unidirectional, the formation of RWs can be explained as a result of wave superposition   \citep{longuet1974breaking,fedele2016real,mcallister2019laboratory,hafner2021real} or modulation instability (MI) \citep{benjamin1967disintegration,zakharov1968stability,tulin1999laboratory,chabchoub2011rogue,bonnefoy2016observation}. Both focusing mechanisms are equally important depending on the wave conditions at play  \citep{dudley2019rogue,waseda2020nonlinear}. 
The nonlinear mechanism in form of MI, along with its manifestation in complex sea states  \citep{tulin1996breaking,waseda2009evolution,onorato2010freak,gramstad2018modulational,toffoli2024observations}, has been extensively studied as a key mechanism for wave group focusing. However, the predominance of quasi-four wave resonant interactions for irregular ocean waves in crossing seas with strong directional spreading is considered to be less evident compared to unidirectional wave field counterparts due to the violation of critical assumptions such as unidirectionality and  narrowband spectral conditions  \citep{janssen2003nonlinear,mori2011estimation,fedele2016real,tang2021unidirection,hafner2023machine}. On the other hand, a recent experimental observation of nonlinear focusing dynamics in standing water waves  \citep{he2022experimental} has shown that MI could still lead to notable amplifications in wave heights for such states. Standing waves can be indeed considered as a simplified specific type of crossing sea state with an aperture angle of $180$ degrees. These findings further illustrate the complex nature of nonlinear wave interactions in complex configurations.

Furthermore and in contrast to the dispersive focusing mechanism in directional wave fields, numerical studies have postulated an increased likelihood of RW formation in coupled two-wave systems by considering weak nonlinearity in the modelling of crossing seas, with the limitations that both wave fields have the same peak frequency and are narrowband \citep{gronlund2009evolution,liu2022statistical}. 
Notably, \cite{liu2022statistical} made a successful attempt to study the crossing RW shape under varying crossing angles and spectral shapes. This approach goes beyond traditional RW investigations, which mainly focus on spectral evolution, exceedance probability distributions, and kurtosis progression. The latter work also highlights that the shape of freak waves is more influenced by the crossing angle between wave components rather than the frequency or directional spectral bandwidth.
The coupled nonlinear Schrödinger equation (CNLS), according to \cite{okamura1984instabilities,onorato2006modulational}, and its higher-order forms \citep{gramstad2011fourth,gramstad2018modulational} are a commonly used frameworks for describing the dynamics of wave envelope interactions in crossing seas  \citep{cavaleri2012rogue}. Indeed, the directional (2D+1) NLS and CNLS frameworks have become essential in understanding the fundamental wave hydrodynamics together with the emergence of localized and directional wave patterns, such as RWs \citep{chabchoub2019directional,steer2019experimental,he2022experimental}. However, it still remains unknown whether the weakly nonlinear wave evolution equations can sufficiently predict the occurrence of all possible RWs, which can occur also in crossing seas. For example, theoretical studies such as in  \citep{guo2020two} based on the Davey-Stewartson (DS) equation  \citep{davey1974three} underline that directional perturbation can trigger strong localizations in time and directional space. Fully nonlinear numerical simulations also predict the spanwise wave instability with some success  \citep{fructus2005dynamics}. These extreme events cannot obviously be predicted by the classical unidirectional NLS framework with their famed breather solutions \citep{akhmediev1985generation,peregrine1983water,chabchoub2011rogue,tikan2022prediction}. On the other hand, the directional NLS can be modified to accommodate exact solutions of the unidirectional and integrable NLS \citep{saffman1978stability,chabchoub2019directional,waseda2021directional} and can describe the dynamics of modulationally unstable short-crested waves. 

To address the remaining key questions as discussed above, our numerical study, which is based on the fully nonlinear numerical framework developed by \cite{wang_modeling_2021}, reveals the existence of a novel type of nonlinear and fully localized RWs, i.e., extreme localized waves in directional space and time, which are distinct in their lifetime from the cases generated as a result of wave overlap. 
The procedure is initiated by accounting for the wave potential's slow variation due to wave nonlinearity in the crossing sea state during the wave data analysis while the crossing New Wave Theory \citep{taylor2004newwave} is adopted as the interference model dynamics for two generated JONSWAP sea states. In this context, we introduce a lifetime parameter we refer to as lifespan $t_{\textnormal LS}$ of a RW event encompassing several consecutive RWs, which was adopted for similar purpose in previous works  \citep{chabchoub2012experimental,kokorina2019lifetimes} and as will be further elaborated upon in detail in the manuscript. Based on the above, we reveal that fully localized RW elevations and their characteristic directional pattern are strongly correlated with their lifespans $t_{\textnormal LS}$. The longer $t_{\textnormal LS}$ of an extreme wave event, the more it differs from a large-amplitude wave created by the interference principle \citep{taylor2004newwave,mathis2015caustics,birkholz2016ocean,fedele2016real}. In the same time, the wave energy exhibits a dual bimodal frequency evolution trend \citep{osborne2017properties,toffoli2010development} in the spreading directional spectrum. Differently than \citep{liu2022statistical}, we do not consider several crossing angles, but rather focus on one particular angle and vary the peakedness parameter of the JONSWAP wave field realizations, i.e. the energy distribution of each of the colliding wave field, and classify the RW events based on their longevity. Finally, we identify a new type of fully localized RW structures characterized by its unique long lifespan $t_{\textnormal{LS}}$ and \textit{full-spatial localization}, i.e., in the mean wave direction, transverse direction, time, and considering the emergence of wave focusing recurrence. Despite having similar features to breathers, such dynamics cannot be predicted by the classical NLS or MI formalism that are applicable only for unidirectional waves. 

\section{Numerical Methods}
Our study embraces a fully and a weakly nonlinear numerical scheme, which will be described in detail below, together with details on the cross wave field initialization. 
\subsection{The Enhanced Spectral Boundary Integral Wave Model: Configuration, Verification, and Validation}

The Cartesian coordinate system is adopted here with \(\bm{x} = (x,y)\) being horizontal and \(z\) being vertical coordinates. The still deep-water level is at \(z=0\). Unless otherwise specified, the variables are being non-dimensionalized, i.e., the distance (\(\bm{x}\), \(z\), and time \(t\) are multiplied by the peak wavenumber \(k_p\) and angular frequency \(\omega_p = \sqrt{gk_p}\), respectively.

The potential flow theory assumes that the fluid is inviscid and irrotational, leading to velocities written as gradients of velocity potential $\phi$, rescaled by \(\sqrt{k_p^3/g}\). The primary advantage of using the velocity potential is that it is a scalar quantity. Therefore, the number of unknowns is reduced compared to the Euler or Navier-Stokes equations, as the velocity vectors can be obtained directly by calculating the gradient of the velocity potential.  

The free surface boundary conditions for the potential flow wave theory consists of those on the water free surface \(z=\eta(\bm{x},t)\): 
 \begin{equation}
 \frac{\partial \eta}{\partial{t} } + \nabla \phi \cdot \nabla \eta - \frac{\partial \phi}{\partial z}  = 0,
 \label{eq:kbc}
 \end{equation}
\begin{equation}
 \frac{\partial \phi}{\partial{t} } + \frac{1}{2}\left[ \left|\nabla \phi \right|^2 + \left(\frac{\partial \phi}{\partial z} \right) ^2 \right] + \eta = 0,
 \label{eq:dbc}
 \end{equation}
 
 \noindent where \(\nabla=(\partial_x,\partial_y)\) is the horizontal gradient operator.

Eqs.(\ref{eq:kbc}) and (\ref{eq:dbc}) are identical to the canonical pair derivable from the Hamiltonian water wave system \citep{zakharov1968stability}. They can be rewritten as a skew-symmetrical form \citep{fructus2005}:
\begin{equation}
\frac{\partial \Psi}{\partial t} + \mathcal{A}\Psi = \mathcal{N},
\label{eq:prognostic}
\end{equation}

\noindent where:
\begin{eqnarray}
\Psi = \begin{pmatrix}
k \mathcal{F}\left\{ \eta \right\} \\
k\omega \mathcal{F}\left\{ \widetilde{\phi} \right\} \\
\end{pmatrix},\quad
\mathcal{A} = \begin{bmatrix}
0 & - \omega \\
\omega & 0 \\
\end{bmatrix}, \quad
\mathcal{N} = \begin{pmatrix}
k\left( \mathcal{F}\left\{ V \right\} - k\tanh{(kh)} \mathcal{F}\left\{ \widetilde{\phi} \right\} \right) \\
\frac{k\omega}{2} \mathcal{F}\left\{ \frac{\left( V + \nabla\eta \cdot \nabla\widetilde{\phi} \right)^{2}}{1 + \left| \nabla\eta \right|^{2}} - \left| \nabla\widetilde{\phi} \right|^{2} \right\} \\
\end{pmatrix},
\end{eqnarray}

\noindent and \(\widetilde{\phi}=\phi(\bm{x},z=\eta,t)\) denotes the velocity potential on free surface, \(V = \sqrt{1 + |\nabla\eta|^{2}}\partial\phi/\partial{n}\) is the vertical velocity of the surface elevation, while \(\mathcal{F}\{\psi\}\) is the Fourier transform defined as:
\begin{equation}
\mathcal{F}\{\psi\} = \iint_{-\infty}^{\infty} \psi e^{-\mathrm{i} \bm{k} \cdot \bm{x}} \mathrm{d} \bm{x},
\end{equation}
with \(\mathcal{F}^{-1}\{\psi\}\) being its inverse transform and \(\mathrm{i}=\sqrt{-1}\). The Fourier transform is implemented numerically by using the Fast Fourier Transform (FFT). 

Equation (\ref{eq:prognostic}) can be further reformulated as:
\begin{eqnarray}
\Psi\left( t \right) = e^{- \mathcal{A}\left( t - t_{0} \right)} \left[   \Psi\left( t_{0} \right) + \int_{t_{0}}^{t}{e^{\mathcal{A}\left( t - t_{0} \right)}\mathcal{N} \mathrm{d} t} \right], 
\end{eqnarray}
and can be used as the prognostic equation for updating unknowns \(\eta\) and \(\widetilde{\phi}\) in time with the integration term evaluated by using a six-stage fifth-order Runge-Kutta method with embedded fourth-order solution \citep{clamond2007note,wang2015}. In general, to keep the difference below 1\%, the time step size is automatically adjusted to about 1/20 peak wave period. For large spatio-temporal simulations of strongly nonlinear waves, a tolerance of 0.1\% is selected corresponding to a time step size of about 1/50 peak wave period, which applies to the crossing sea simulations in the present study.

In order to  update the solutions \((\eta,\widetilde{\phi})\), the velocity \(V\) needs to be diagnosed by solving the boundary integral equation of Green's theorem. 
A successive approximation approach can be adopted, and the total vertical velocity is expressed as \(V = \sum V_m\), where \(m\) represents the order of the nonlinearity $\mathcal{O}(\varepsilon^m)$, where the expansion parameter $\varepsilon$ denotes the wave steepness. For simplicity, the recurrence formula for estimating $V_m$ in the fully nonlinear Enhanced Spectral Boundary Integral (ESBI) wave model in deep-water starts with:

\begin{equation}
\mathcal{F}\left\{ V_{m} \right\} = k \mathcal{F}\left\{ \widetilde{\phi} \right\} ,
\end{equation}

\noindent and then the remaining velocities in Fourier domain will be calculated by:

\begin{eqnarray}
\mathcal{F}\left\{ V_{m} \right\} &=&  \sum_{j=1}^{m-1}   - \frac{k^j}{j!}\mathcal{F}\left\{ \eta^jV_{m-j} \right\} - \frac{k^{m-2}}{(m-1)!} \mathrm{i}\bm{k} \cdot \mathcal{F}\left\{ \eta^{m-1}\nabla\widetilde{\phi}  \right\} ,    
\end{eqnarray}
\noindent for $m \ge 2$. 

In this pseudo-spectrum method, the \(2/(m+1)\)-rule is used here for anti-aliasing treatment, which is equivalent to the zero-padding method \citep{canuto1987spectral}. We emphasize that a smoothing technique is not required here, and the present model is very stable for the cases without appearance of breaking waves. The model has been comprehensively verified and validated for simulating a variety of highly-nonlinear wave phenomenon,  crossing seas, and laboratory experiments \citep{wang2018fully, wang_modeling_2021,wang2023enhanced}.

Before initiating our study, we recall the commonly used definition of a RW, namely having a crest height $\eta_c$, exceeding at least 1.25 times the significant wave height $H_s$, i.e., $\eta_c>1.25 H_s$. In a Gaussian sea state $H_s$ is approximated of being four times the standard deviation of the entire water surface elevation. This threshold is based on previous studies and represents a significant deviation from the mean wave height  \citep{kharif2008rogue,gramstad2018modulational,mori2023science}. In this context, we define $t_{\textnormal{LS}}$ as the duration or the lifetime of a series of observed sequence of extreme waves belonging to the same RW event. This will be clarified and discussed later on in the manuscript. 

The numerical setup is described as next. The computational domain of the simulations covers 40 \(\times\) 40 peak wavelengths, and is resolved into 1024 \(\times\) 512 collocation points in $x$ (along-wave) and $y$ (cross-wave) directions, respectively. 
While the selected domain size and resolution in space ensure that the Fourier modes up to seven and three times peak wavenumber in the $x$- and $y$- directions, respectively, are aliasing-free. 
The reference sea state and wave surface condition is based on the JONSWAP spectrum \citep{hasselmann1973measurements} with different peakedness factors \(\gamma = 1,\ 2,\ 3,\ 4,\ 6,\) and $9$ with crossing wave field of steepness \(k_pH_s = 0.28\).  These two wave systems with same peak frequency $f_p$, peakedness parameter $\gamma$, significant wave height $H_s$, and random phases  cross-interact at an aperture angle of $40$ degrees, which corresponds to the most hazardous angle  leading to the highest probability distribution tail \citep{toffoli2011extreme,cavaleri2012rogue,bitner2014occurrence}. Several wave probes are deployed every four peak wavelengths along the center of the domain in $x$-direction. The wave generation zone is deployed along \(x=0\) and absorbed at a distance near the other end. Each simulation lasts for 1000 peak periods $T_p$ (equivalent to a typical three hours sea state), and four realizations are performed. 

\begin{figure}
\centering
   \includegraphics[width=1\textwidth]{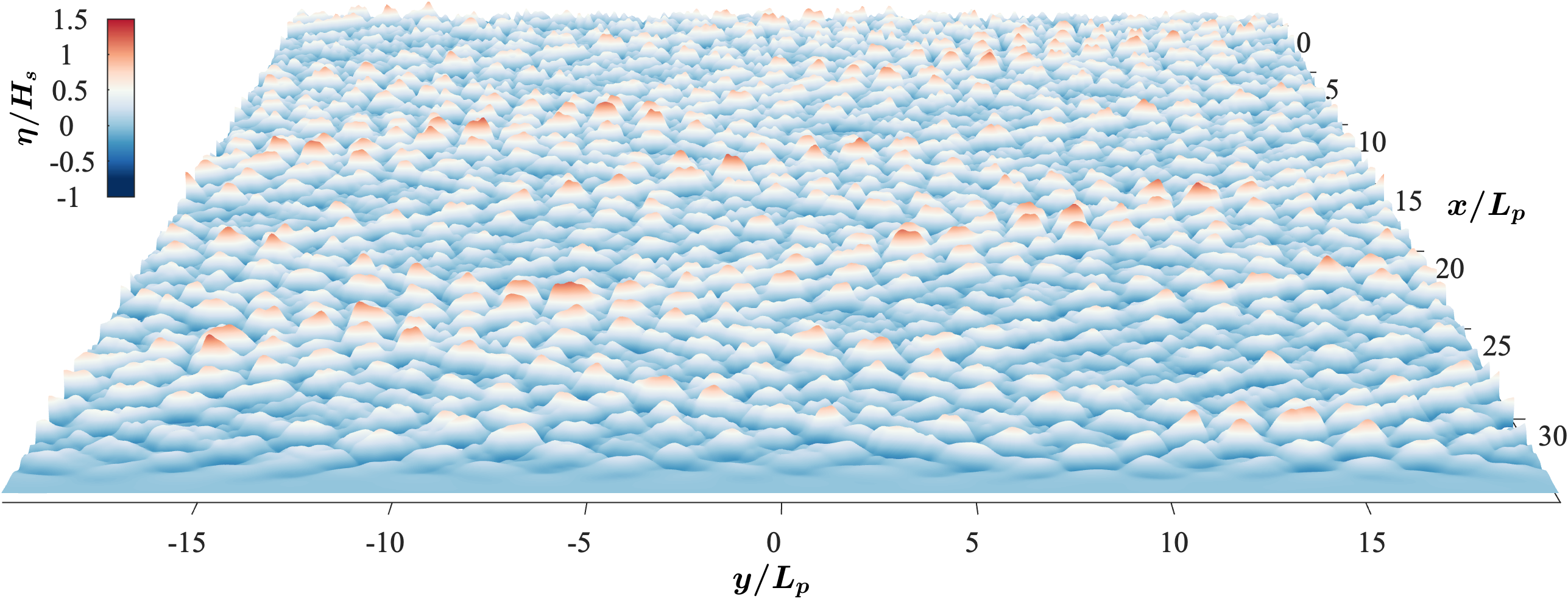}
\caption{An exemplified snapshot example of a simulated crossing sea surface elevation of steepness $k_pH_s=0.28$ for a cross-interfering JONSWAP wave field with peakedness $\gamma=6$. 
}
\label{fig:cross-surf}
\end{figure}

A snapshot of the simulated cross free surface as described above is shown in Figure \ref{fig:cross-surf}. Several large amplitude wave groups can be observed in directional space at a particular instant of time. 
We would like to emphasize that wave breaking is inevitable under such wave conditions. To stabilize the simulations, the low-pass filter is employed to suppress the breaking \citep{xiao2013rogue} . This filter is shown to well represent the energy dissipation quantitatively over a broad range of wave steepness, breaker types and directional spreading. We refer to \citep{wang2021modeling} for a direct comparison of the model simulations with laboratory experiments, involving the evolution of kurtosis and wave crest exceedance probability trends.

\subsection{The Hydrodynamic Coupled Nonlinear Schrödinger Equation}
 
The purpose for introducing the CNLS is two-fold. It is adopted to compare the strong RW localizations, as obtained from the fully nonlinear ESBI simulations, to a weakly nonlinear framework which applies for cross sea modelling \citep{cavaleri2012rogue}, and to test if a weakly nonlinear wave framework is sufficient to characterize all measured RWs observed in the ESBI simulations. We carried out benchmarking numerical simulations by means of the CNLS, as previously reported and parametrized for water waves \citep{okamura1984instabilities,onorato2006modulational} and mentioned in the introductory Section \ref{section-intro}. The two-wave deep-water coupled framework writes:
\begin{equation}
\begin{split}
    &\frac{\partial u_1}{\partial t}+c_x\frac{\partial u_1}{\partial x}+c_y\frac{\partial u_1}{\partial y}-\mathrm{i}\alpha \frac{\partial^2 u_1}{\partial x^2}-\mathrm{i}\beta \frac{\partial^2 u_1}{\partial y^2}+\mathrm{i}\gamma \frac{\partial^2 u_1}{\partial x \partial y}+\mathrm{i}\left(\xi |u_1|^2 + 2\zeta |u_2|^2 \right) u_1=0,\\
    &\frac{\partial u_2}{\partial t}+c_x\frac{\partial u_2}{\partial x}-c_y\frac{\partial u_2}{\partial y}-\mathrm{i}\alpha \frac{\partial^2 u_2}{\partial x^2}-\mathrm{i}\beta \frac{\partial^2 u_2}{\partial y^2}-\mathrm{i}\gamma \frac{\partial^2 u_2}{\partial x \partial y}+\mathrm{i}\left(\xi |u_2|^2 + 2\zeta |u_1|^2 \right) u_2=0,
    \end{split}
    \label{equation-CNLSE}
\end{equation}
where the coefficients write are defined as:
\begin{equation}
\begin{split}
    c_x=\frac{\omega}{2\kappa^2}k,\ c_y=\frac{\omega}{2\kappa^2}&l,\ \alpha=\frac{\omega}{8\kappa^4}(2l^2-k^2),\ \beta=\frac{\omega}{8\kappa^4}(2k^2-l^2),\\
    \xi=\frac{1}{2}\omega\kappa^2,\ \zeta=\frac{\omega}{2\kappa}&\frac{k^5-k^3l^2-3kl^4-2k^4\kappa+2k^2l^2\kappa+2l^4\kappa}{(k-2\kappa)\kappa},\\
    &k=\kappa \cos{\theta},\ l=\kappa \sin{\theta}.
\end{split}
\end{equation}
Here, $u_1(x,y,t)$ and $u_2(x,y,t)$ are two crossing complex wave envelopes with wavenumbers of $k_1=(k,l)$ and $k_2=(k,-l)$, respectively, $\theta$ is the crossing angle, the angular frequency $\omega$ and the modulus of their wavenumbers $\kappa$ obey the deep-water dispersion relation $\kappa=\omega^2/g$, where $g$ is the gravitational acceleration. Note that when $\theta=0$, $u_2$ is inactive, and therefore, Equation (\ref{equation-CNLSE}) is uncoupled and naturally reduces to the classical NLS \citep{zakharov1968stability}. The first-order approximation of a two-wave-field crossing elevation $\eta(x,y,t)$ is given by:
\begin{equation}
    \eta(x,y,t)=\frac{1}{2}\left(u_1(x,y,t)e^{\mathrm{i}(kx+ly-\omega t)} +u_2(x,y,t)e^{\mathrm{i}(kx-ly+\omega t)}+c.c.\right),
\end{equation}
where $c.c.$ denotes the complex conjugation. To ensure highest numerical accuracy, fourth-order Runge-Kutta and pseudospectral methods \citep{yang2010nonlinear} are adopted to advance Equation \ref{equation-CNLSE} in time. The two equations are coupled in a staggered manner, as already adopted by \cite{he2022experimental} to validate laboratory observations. The corresponding numerical results will be reported and compared to the fully nonlinear ESBI results in Subsection \ref{section-comparison with nlse}.

\section{Lifespan Analysis of Emerged Rogue Waves and Categorization}
We begin by reasonably assuming that crossing RWs develop along the mean wave direction \citep{onorato2006modulational}, and trivially define that two RWs are considered part of one "event" if their adjacent distance does not exceed two wavelengths in the mean wave direction and 0.8 wavelengths in the perpendicular direction. In addition to that, the latter events must occur within a time interval not exceeding five periods. Thus, the \textit{lifespan} of an independent event $t_{\textnormal{LS}}$ can be defined as: 
\begin{equation}
    t_{\textnormal{LS}}=t_\textnormal{RW,end}-t_\textnormal{RW,ini},
\end{equation}
where $t_\textnormal{RW,ini}$ and $t_\textnormal{RW,end}$ are the occurrence time of the first and last identified RW in an identified extreme localization event. 
In particular, we refer to the events with $t_{\textnormal{LS}}/T_p=1$ as those short-lived ones with a lifespan less than one peak wave period.

\begin{figure}
  \centering
  \includegraphics[width=0.9\textwidth]{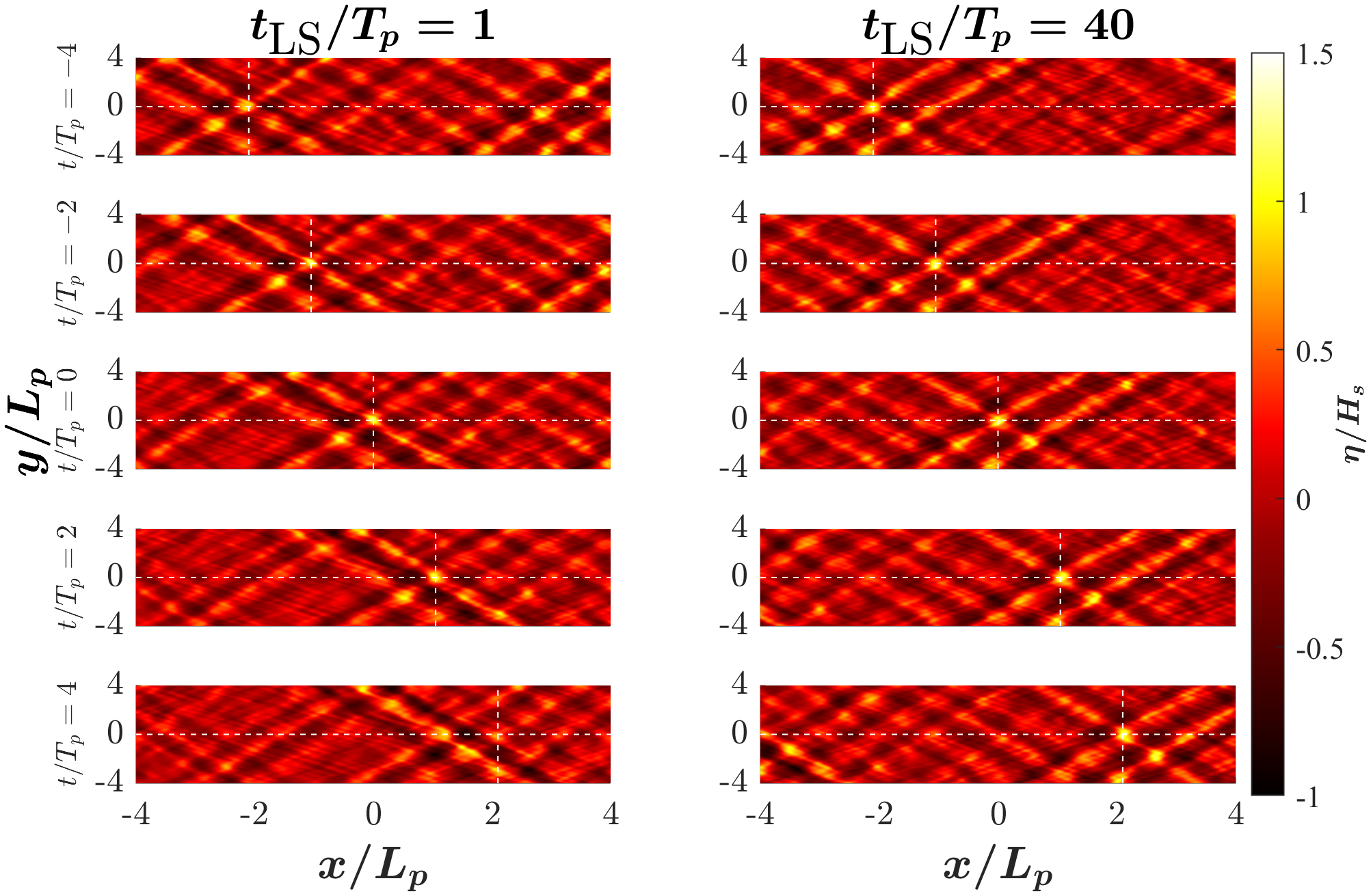}
 \caption{Comparison of two exemplary cross-RW events with different lifespans $t_{\textnormal{LS}}$, {evolving from the top to bottom in each column}. The mean-direction wave field evolution is plotted every two peak wave periods $T_p$. When $t/T_p=0$, the current RWs reach their peak. The left five subplots show a short-lifespan RW event with $t_{\textnormal{LS}}=1T_p$ only. The subplots on the right show a long-lifespan RW event with $t_{\textnormal{LS}}=40T_p$. 
 The orthogonal white dashed lines indicate the location of the compared RW events relative to the reference center point.
 Both cases are extracted from the numerical JONSWAP sea state simulations for $k_pH_s=0.28$ as already described in Figure \ref{fig:cross-surf}.} \label{figure tracking peak locations}
 \end{figure}

From Figure \ref{figure tracking peak locations}, two representative RWs with different $t_{\textnormal{LS}}$ are compared, suggesting that RWs appearing at different time scale lengths along the mean wave direction can be observed within the same crossing wave field. To simplify the further analysis, we will treat such cases as \textit{independent events} while the first class of such events being conjectured as a result of wave interference, due to the very short focusing time. 

 \subsection{Spatio-Temporal Evolution of Long-Lived Directional Coherent RW Structures}
\begin{figure}
\centerline{\includegraphics[width=0.9\textwidth]{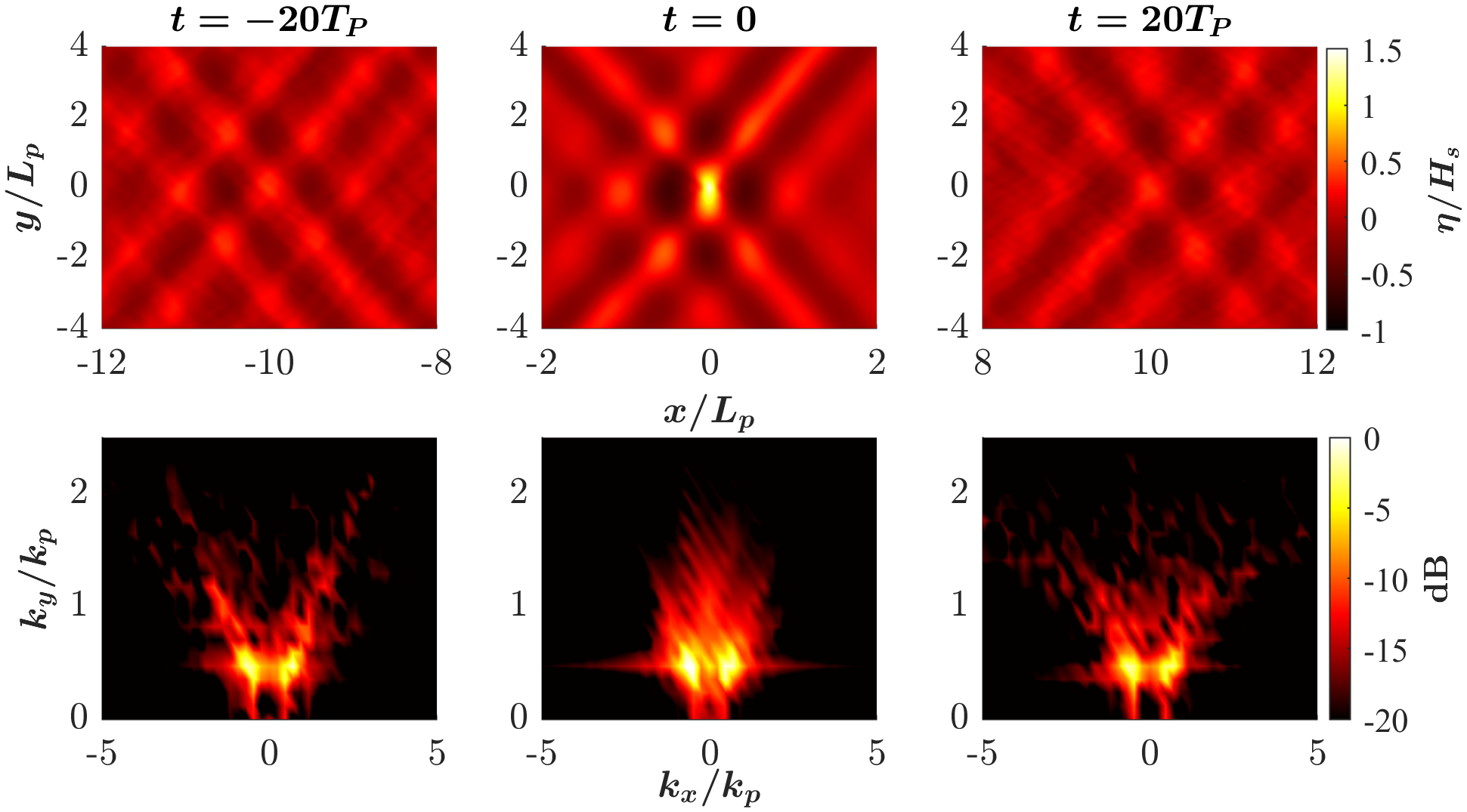}}
\caption{Spatio-temporal evolution of the mean 3D RW events with long-lifespan ($t_{\textnormal{LS}}>35 T_p$) and JONSWAP peadkedness factor $\gamma=3$ at three different stages of extreme focusing evolution: $20T_p$ before the peak, at the peak ($t=0$), and $20T_p$ after the peak, which correspond to the left, middle, and right plots, respectively. Upper plots: The mean elevation field. Lower plots: Corresponding mean 
 wave energy spectra. \label{figure eta and spectrum evolution}}
\end{figure}

To further examine whether the observed long-lived localized wave structures experience both growth and decay, and thus indeed obey the definition of a RW, we depict in Figure \ref{figure eta and spectrum evolution} the mean spatio-temporal evolution of all long-lifespan RW elevation fields followed by the corresponding mean directional spectra. 

In the current figure, we can observe a strong wave amplitude focusing followed by a decay in both $x$ and $y$-directions, as well as the broadening and narrowing of the directional spectrum. From the current observation, we show that a clearly detectable {\it long-lived breathing-type} process can occur in directional seas, as suggested by our fully nonlinear framework simulations.  Thus, these observed directional and particular RW structures can be a result of nonlinear interactions, which yield to a \textit{full} localization in time and directional space. To the best of our knowledge such pulsating wave phenomenon with such longevity features was so far not reported and not discussed by means of fully nonlinear numerical simulations.

To further characterize and distinguish these RW events in such irregular cross sea states, it is important to extract key statistical characteristics, which will be analyzed and discussed as next. 

\subsection{Statistical Analysis of RW Events in Crossing Seas}
\begin{figure}
\centerline{\includegraphics[width=1\textwidth]{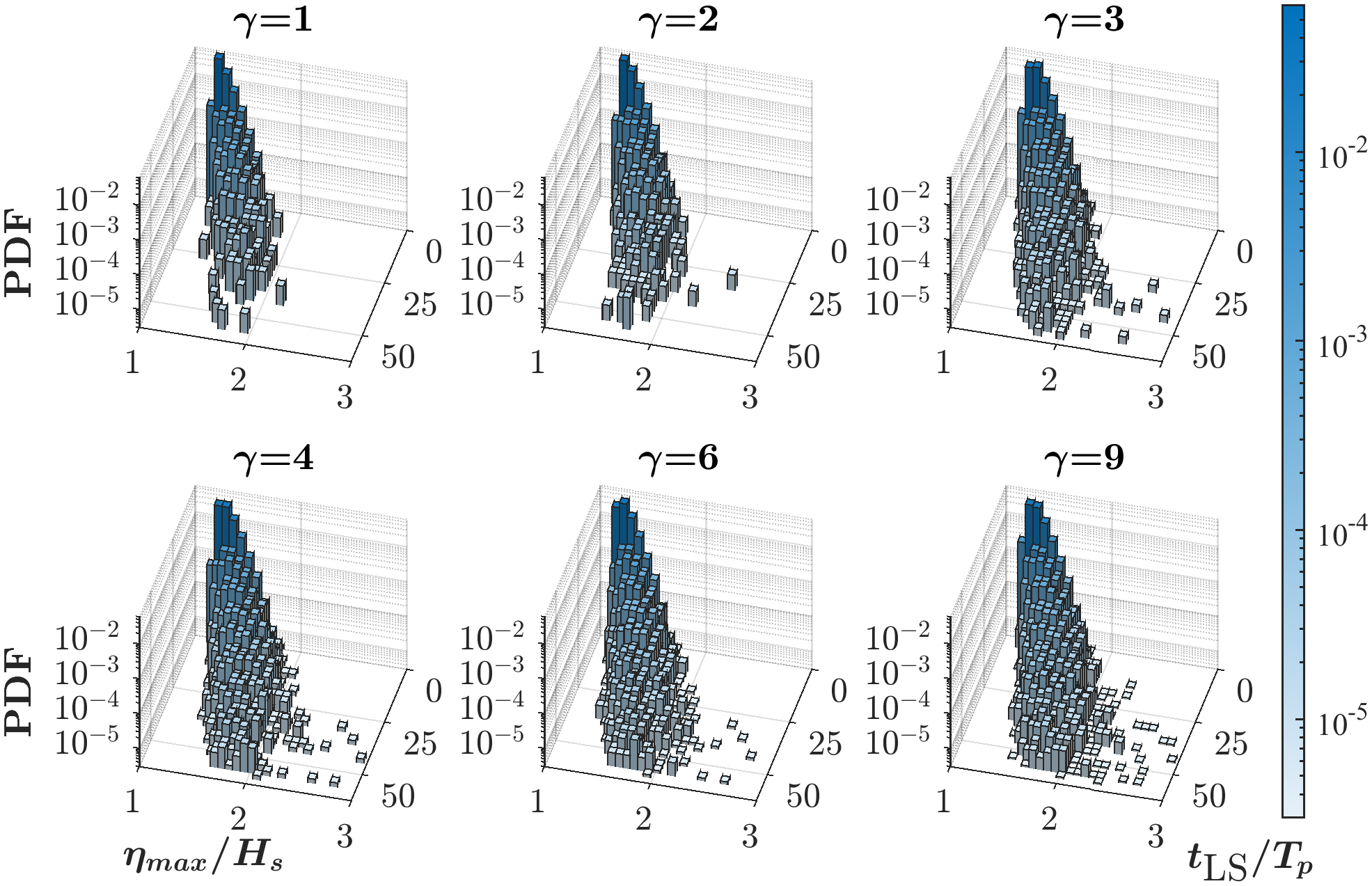}}
\caption{Statistical PDF results of the correlation between the amplification factors, defined as $\eta_{\textnormal{max}}/H_s$, their lifespans $t_{\textnormal{LS}}$ as RWs along the mean wave directions, and the probability density function while considering different JONSWAP spectral peakedness parameter $\gamma$ values. \label{figure tracking peak locations statistics}}
\end{figure}

In the following, we extract all independent events from the numerical data and calculate their corresponding probability density function (PDF) with a categorization with respect to both, $t_{\textnormal{LS}}$ and maximum wave crest height $\eta_{\textnormal{max}}$ of the extreme events. As shown in Figure \ref{figure tracking peak locations statistics}, the PDFs of all independent events are generated with respect to the maximum amplification factor and the normalized lifespan for all six JONSWAP-$\gamma$ cases with the identical $H_s$ values as mentioned earlier. 
The results indicate that most independent events have a short lifespan, suggesting that wave superposition principle is a dominant focusing mechanism for our modelled cross sea states. That being said, around one in $10^3$ to $10^5$ events exhibits significantly longer lifespan, depending on the JONSWAP peakedness parameter $\gamma$ value considered, with a trend of increasing amplification factor by gradually narrowing the initial energy spectrum. Such an observations indicate the existence of a nonlinear focusing mechanism responsible for such wave amplifications, despite having a low probability. However, the fact that such rogue waves are recurrent, i.e., experience recurrent focusing, these can indeed still pose a significant threat in the ocean.

\begin{figure}
\centerline{\includegraphics[width=1\textwidth]{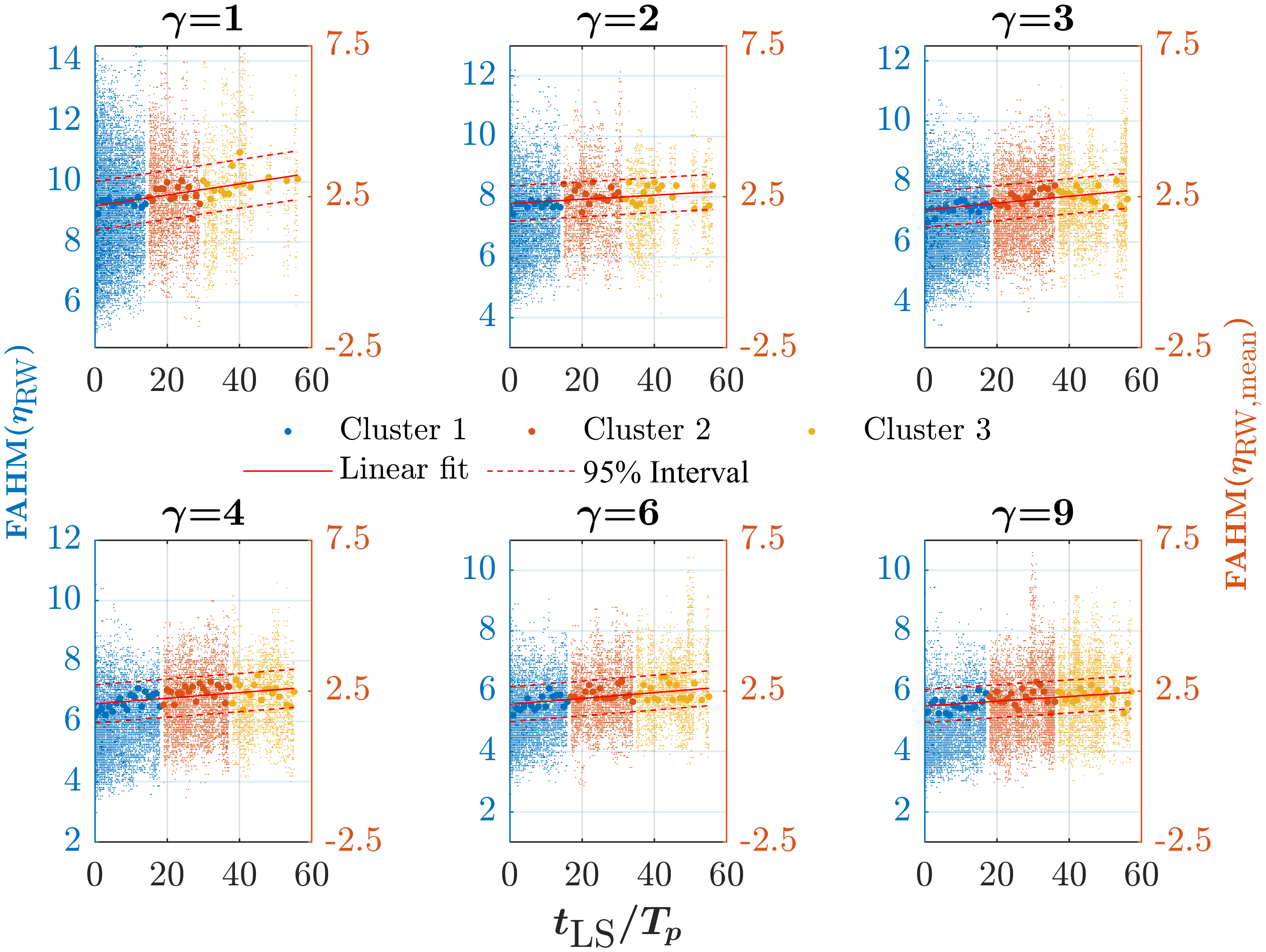}}
\caption{The correlation between the RWs' lifespans $t_{\textnormal{LS}}/T_p$ and their directional FAHM factor. The left $y$-axis represents the FAHM values of all RWs visualized by minor dots, and the right $y$-axis shows the FAHM values of the mean RW elevation fields (instead of the mean FAHM value) depicted by thicker dots. Here, an averaging interval of $t_{\textnormal{LS}}=0.6$ is used. Considering the mean FAHM, the data are categorized into three K-means clusters in different colors. The linear fit and $95\%$ confidence interval are also given for each case.   \label{figure tracking peak locations statistics FAHM}}
\end{figure}

To interpret these nonlinear RW events, as we will call these from now on, we analyze the events according to their lifespans and the spectral directionality measured by the "Full \textit{Area} Half Maximum" (FAHM), see Figure \ref{figure tracking peak locations statistics FAHM}. Since the distributions of the RWs are highly non-uniform along the $t_{\textnormal{LS}}/T_p$ axis, a good linear fit and reasonable clustering can be achieved upon averaging all original RW elevation data along the $t_{\textnormal{LS}}$ axis within every certain interval, in our case chosen to be $t_{\textnormal{LS}}/T_p=0.6$. 

Surprisingly, those nonlinear RW events with longer lifespans show greater FAHM levels, suggesting the occurrence of a long-term broadening of the directional wave spectrum. Here, we highlight the work by  \cite{toffoli2010development}, which numerically observed the development of a similar bimodal pattern in the spectra. {Meanwhile, the analysis by \cite{osborne2017properties} shows that an initial and single JONSWAP spectrum can also become directionally-unstable along the modulation channel}. However, these strongly relevant studies were not directly related to the formation of full-spatial localized and directional RW structures. 
For further ease of the analysis, the RW events under each $\gamma$ case considered in Figure \ref{figure tracking peak locations statistics FAHM} are categorized into three K-means clusters \citep{lloyd1982least,arthur2007k,cremonini2021selection}, by using different colors, each representing a group of independent RW events with different FAHM and lifespan ranges. Note that the clustering is mainly used to group the many RW data observed from the simulations, and is achieved by adopting the K-means++ algorithm as developed by \cite{arthur2007k}, which is more effective than the standard K-means algorithm \citep{lloyd1982least}.

\begin{figure}
\centerline{\includegraphics[width=1\textwidth]{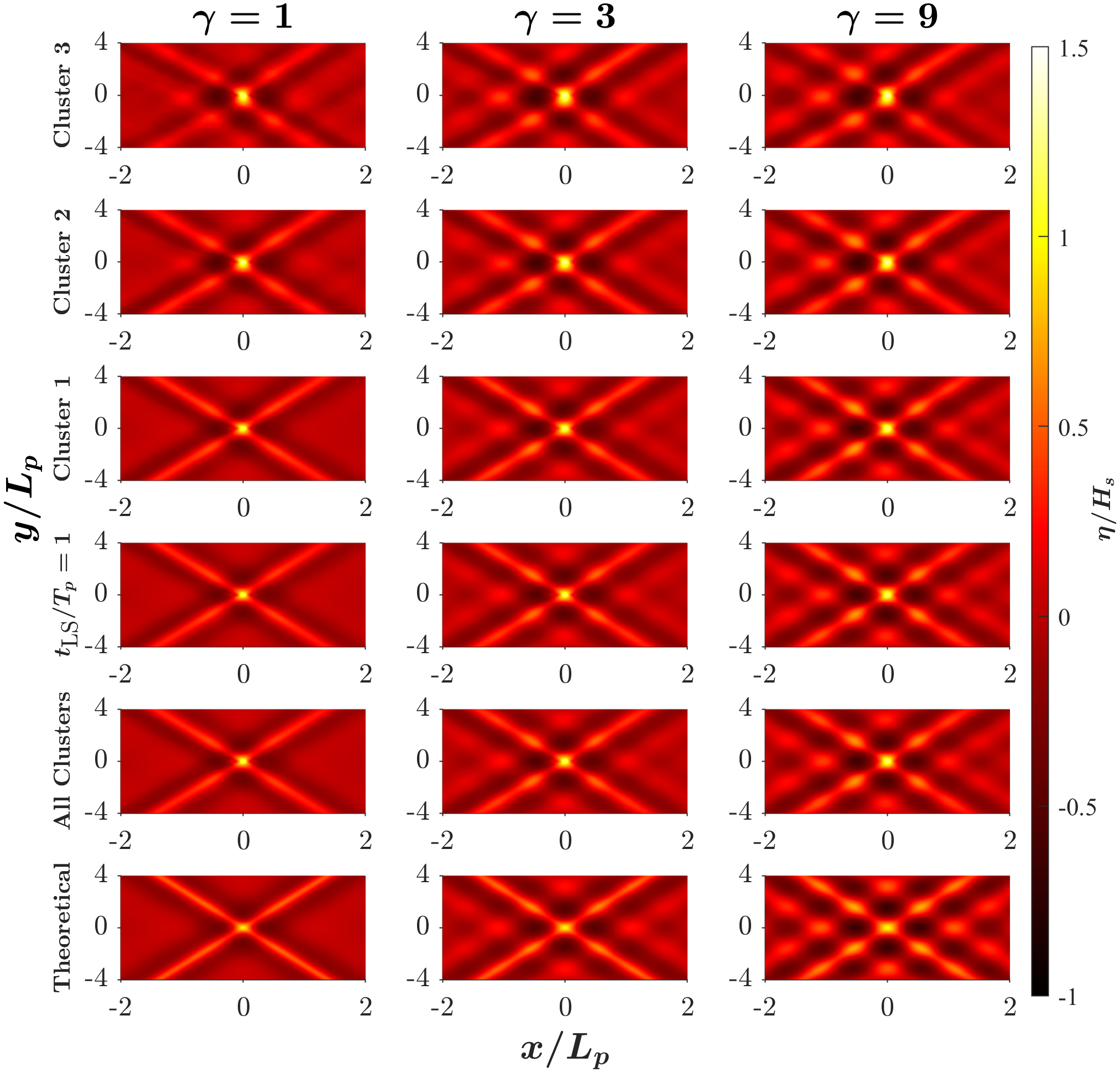}}
\caption{3D averaged RW wave local elevation field, truncated according to the three K-means clusters from Figure \ref{figure tracking peak locations statistics FAHM} (top three rows) according to Figure \ref{figure tracking peak locations statistics FAHM}, and compared with the averaged RW elevation field calculated from all $t_{\textnormal{LS}}=1T_p$ cases corresponding to {one-RW events} (third-bottom row), with all RW events as reference (second-bottom row), and the corresponding second-order New Wave Theory spectrum (bottom row). Only three representative JONSWAP peakedness parameter factors $\gamma$=$1$, $3$, and $9$ are considered in the current figure.  \label{figure tracking peak locations statistics eta shape}}
\end{figure}

Upon averaging the RW events within each cluster and while tracking the directional wave elevation field evolution, we can notice in Figure \ref{figure tracking peak locations statistics eta shape} deviations from the New Wave theory \citep{taylor2004newwave}, with lifetime $t_{\textnormal{LS}}/T_p=1$ (bottom row), increases with the increase of $t_{\textnormal{LS}}$ or the cluster number, especially, for Cluster 3, in which the elevation fields are nearly independent of the initial JONSWAP-peakedness $\gamma$ values and form four dark wave trough holes around the center peak. 

Whether such a coherent structure implies a possible deterministic description of so-called higher-dimensional nonlinear RWs, like it is the case for NLS breathers for unidirectional wave states, needs future attention.  

The shape of these extreme waves is similar to the ones already reported in \citep{liu2022statistical}. Differently to the latter work, we focus our analysis on the longevity of the rogue wave events while considering one collision angle only and varying the JONSWAP peakedness parameter. 

\begin{figure}
\centerline{\includegraphics[width=1\textwidth]{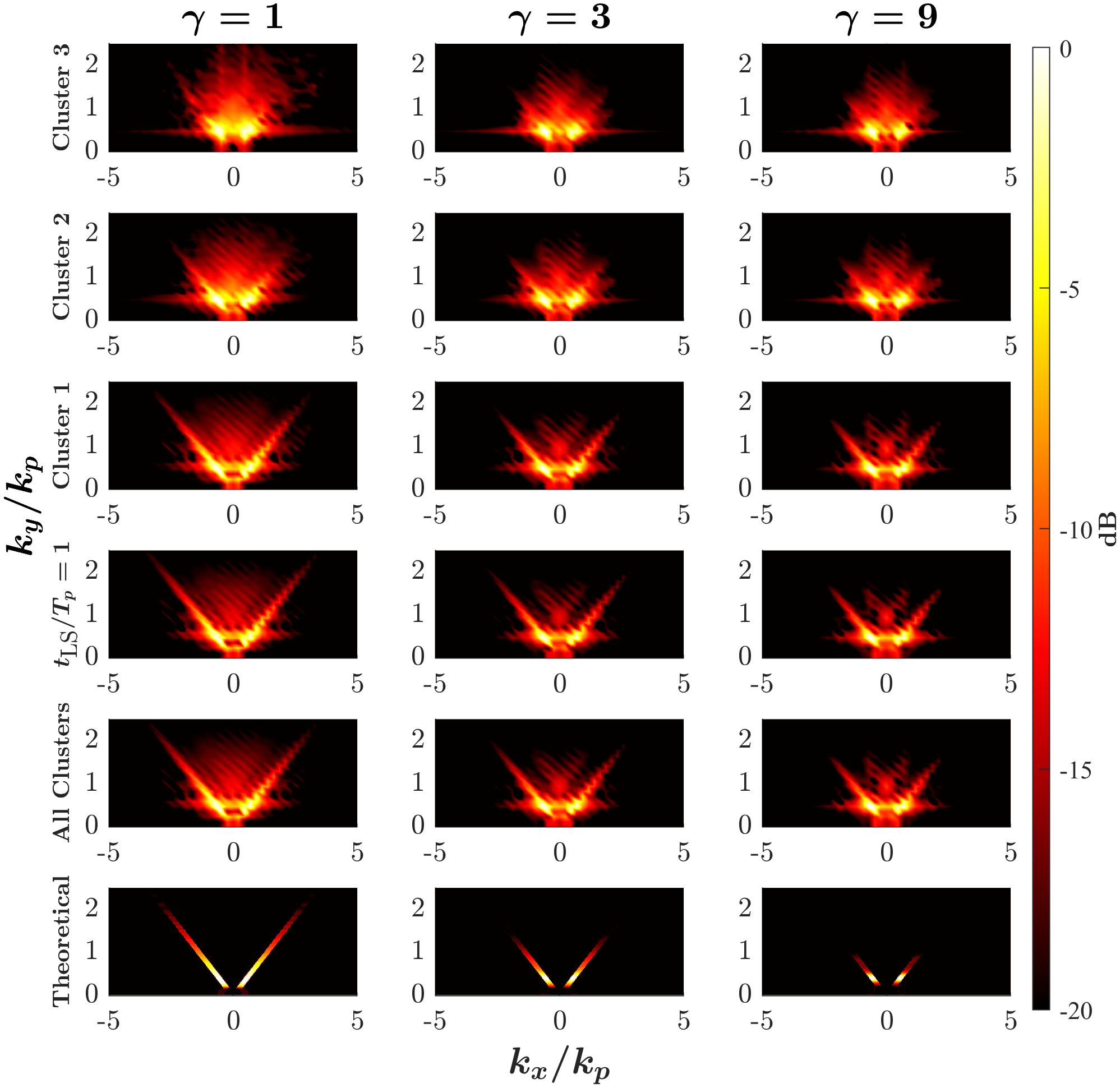}}
\caption{3D averaged RW local spectra corresponding to Figure \ref{figure tracking peak locations statistics eta shape}. }
\label{figure tracking peak locations statistics spectrum}
\end{figure}

Next, we show the corresponding spectra of all cases discussed and shown in Figure \ref{figure tracking peak locations statistics spectrum}, confirming once again the significant broadening of the directional spectrum from short- to long-lived $t_{\textnormal{LS}}$ RW events. Note that with longer $t_{\textnormal{LS}}$, the directional spectrum gradually differs from the wave interference case, which is rather characterized by remaining narrowband in the two wave directions after the extreme wave focusing. {Despite, from both Figures  \ref{figure tracking peak locations statistics eta shape} and  \ref{figure tracking peak locations statistics spectrum}, one can notice that averaging all RW events can easily lead to the neglect of Clusters 2 and 3, which are the most affected by nonlinearity and directional spreading, develop of a dual bimodal structure in the spectrum. It is therefore recommended to consider and adopt an appropriate aggregated approach to analyze the real-world RW data  \citep{hafner2021real}.} 

\subsection{Comparison with the CNLS Framework}\label{section-comparison with nlse}
In order to further understand the role of nonlinearity and particularly the degree and order in the manifestation of these coherent directional large-amplitude waves in a crossing sea setup, we compare the obtained fully nonlinear results with the CNLS framework simulations.

\begin{figure}
\centerline{\includegraphics[width=1\textwidth]{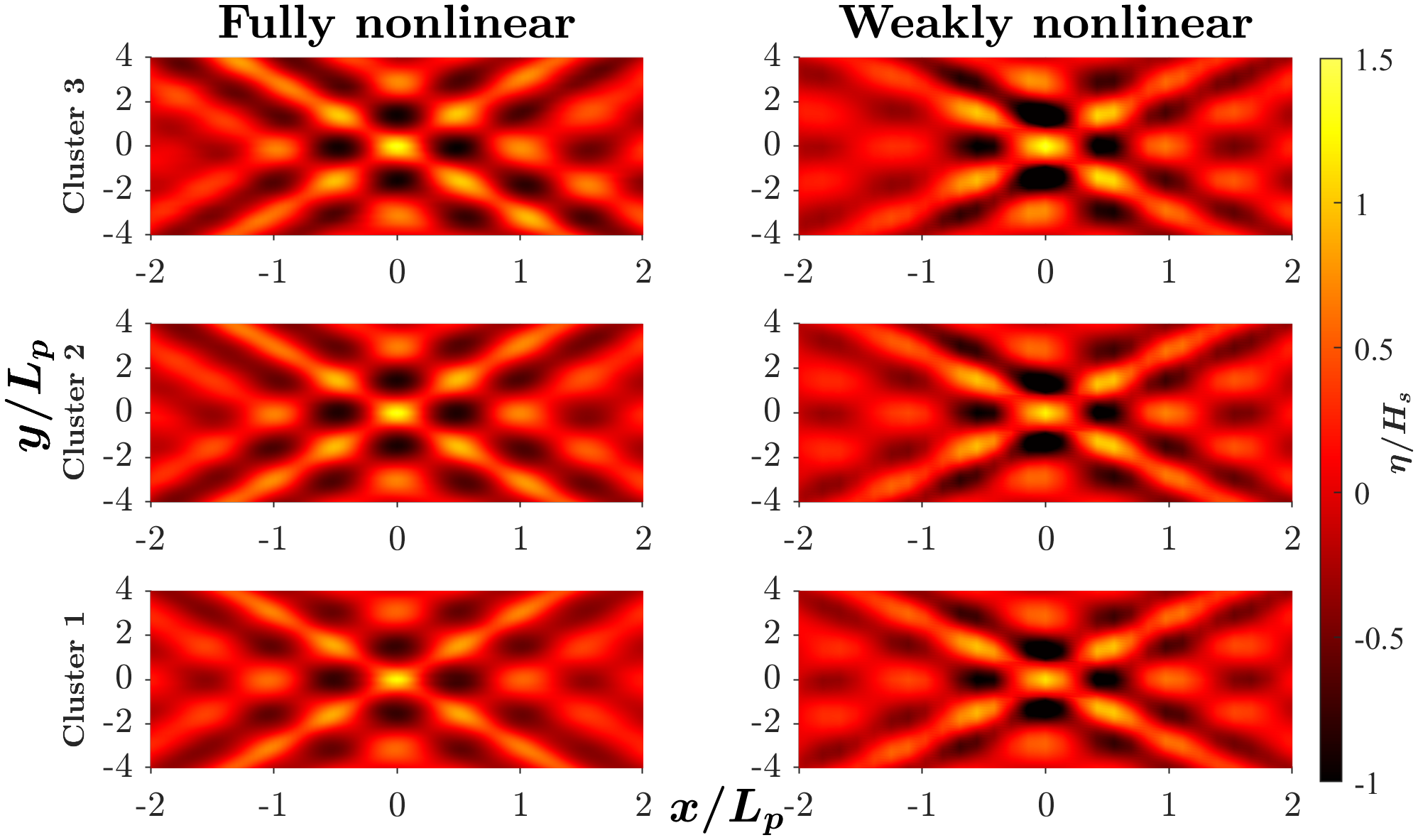}}
\caption{Comparison of directional and averaged RW wave local elevation field simulated by means of the fully nonlinear (ESBI) framework (left panels) and the corresponding CNLS simulations (right panels). The generate JONSWAP wave field has a significant wave height of $H_s=0.015$ m with $\gamma=9$.  \label{figure CNLSE eta}}
\end{figure}

Classified into three K-means clusters, the directional and localized RW elevation results in Figure \ref{figure CNLSE eta} show a trend similar to those in Figure \ref{figure tracking peak locations statistics eta shape} for a low significant wave height $H_s=0.015$ m of the JONSWAP wave field generated in each direction with $\gamma=9$. Such wave height scales have been chosen to motivate future laboratory experiments. 
\begin{figure}
\centerline{\includegraphics[width=1\textwidth]{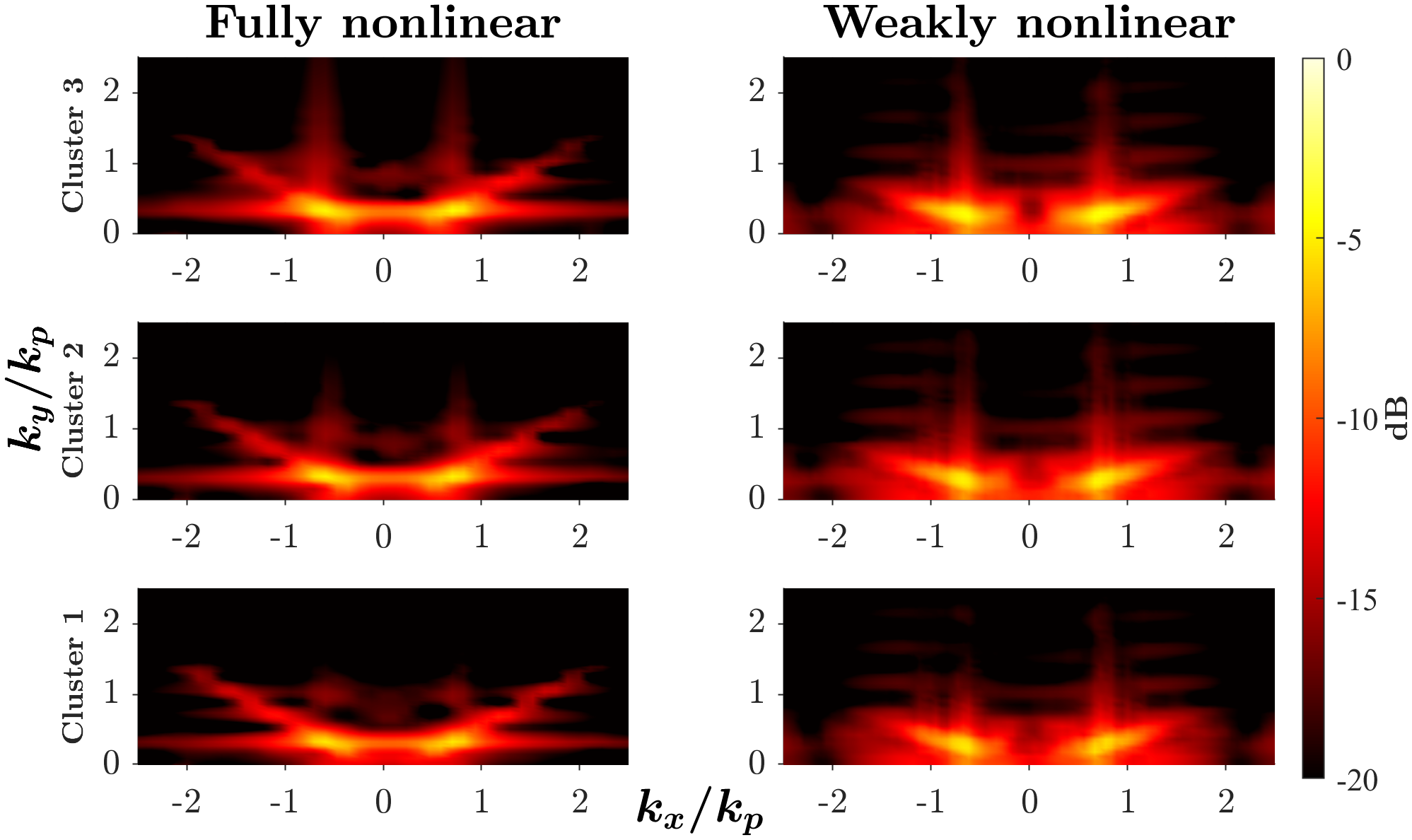}}
\caption{Comparison of the corresponding averaged RW wave local spectra corresponding to Figure \ref{figure CNLSE eta}.    \label{figure CNLSE spectra}}
\end{figure}
Differences in averaged wave envelope shapes are noticeable in Figure \ref{figure CNLSE eta}, particularly for the long-lived extreme waves, which are clumbed in Cluster 2 and 3. The directional spectra in Figure \ref{figure CNLSE spectra} further confirm a typical and expected spectral broadening in both fully and weakly nonlinear frameworks. In fact, the CNLS predicts a broader initial spectrum and reduced dual bimodality trend, compared to the fully nonlinear ESBI simulations.

\begin{figure}
\centerline{\includegraphics[width=1\textwidth]{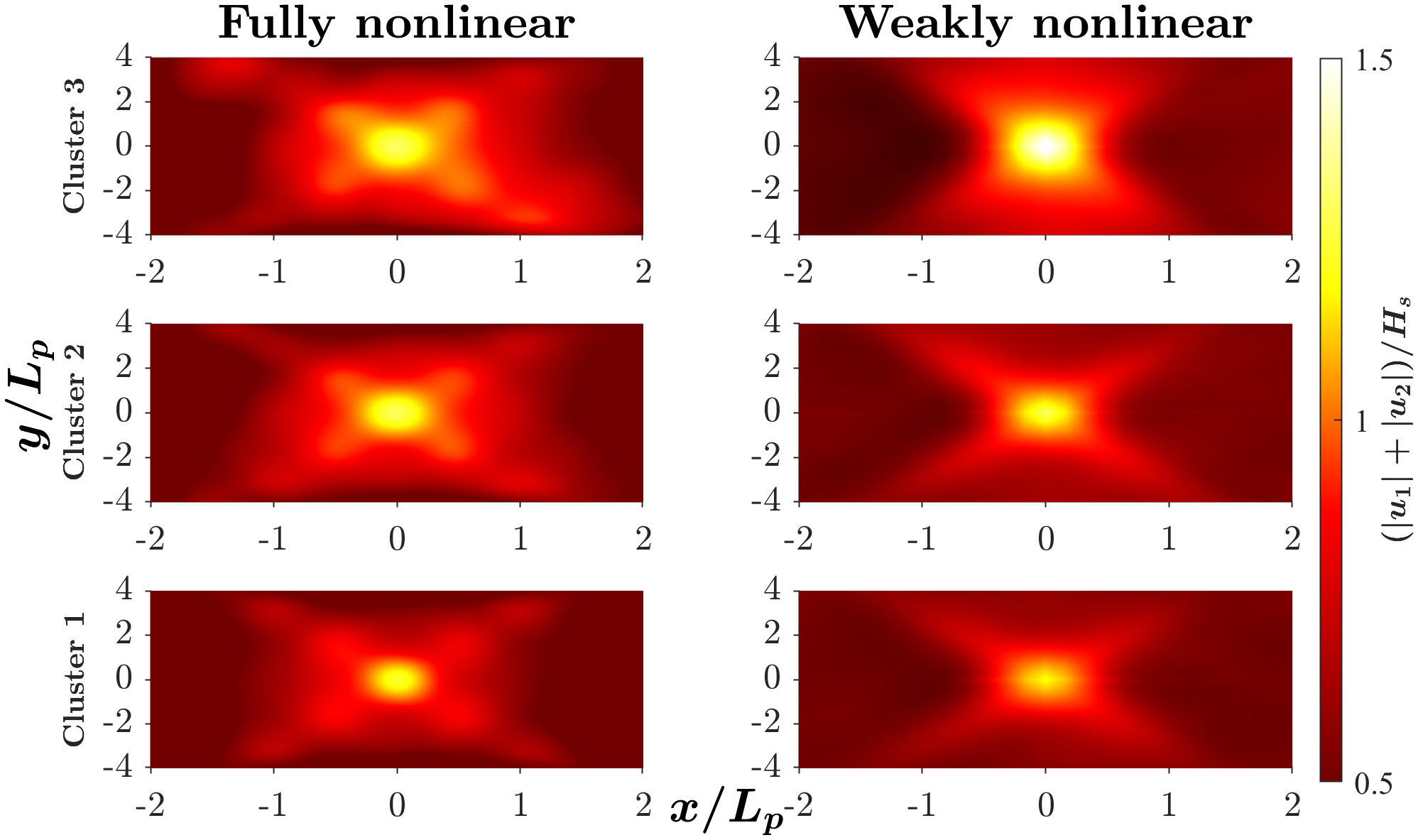}}
\caption{Comparison of the magnified and directional averaged localized maximal envelope peak shapes corresponding to Figure \ref{figure CNLSE eta}.    \label{figure CNLSE envelope}}
\end{figure}

Intriguingly, one can clearly observe the differences in the extreme wave envelope peak, as magnified in Figure \ref{figure CNLSE envelope}. While the Cluster 1 wave envelopes at the bottom of the figure, which appear to be the result of wave overlap, have a quasi-similar pattern, the long-lived extreme events in Cluster 2 and 3 show a completely different and distinct shape when simulated by the fully and weakly nonlinear framework. Such directional localized wave patterns cannot be analytically modelled thus far. A complete and quantitative characterization of these coherent extreme waves requires comprehensive future numerical and experimental explorations.

\section{Discussion and Conclusion}
This paper investigates key physical and statistical properties of RW events in crossing JONSWAP-type sea states using a fully nonlinear ESBI framework \citep{wang_modeling_2021} and a lifespan-based analysis approach \citep{chabchoub2012experimental,kokorina2019lifetimes}. We focus on a specific crossing angle of 40 degrees and analyze RWs under different JONSWAP spectrum bandwidths \citep{toffoli2011extreme,cavaleri2012rogue,bitner2014occurrence}.

Our numerical ESBI results reveal RW events developing along the mean wave direction with lifespans ranging from $1T_p$ while satisfying the RW threshold criterion, up to $40 T_p$ and beyond. For these long-lasting nonlinear RW events, we observe a clear focusing and decay process of the mean directional wave elevation field, along with a perfect recurrence of the mean directional wave spectrum. This suggests that wave superposition and the classical (unidirectional) MI \citep{benjamin1967disintegration,mori2011estimation} are insufficient for the prediction of all extreme wave events in such realistic directional seas.

On the other hand, we further analyze the statistical characteristics of such occurring crossing RW events by varying the JONSWAP peakedness factor $\gamma$ and find that the probability of long-lifespan events increases by increasing $\gamma$, highlighting the potential role of quasi-four wave resonant interactions in one of the two colliding wave beams, which might be not substantially influenced by the other wave field component. Moreover, previous studies point to the significance of spreading effects \citep{toffoli2010development} as well as directional weakly nonlinear effects \citep{osborne2017properties} in the RW generation. The emergence of the localized severe and typical spectral broadening followed by a dual bimodality further supports the role of nonlinear wave interactions in the formation of RWs in cross seas. 

Furthermore, classifying RW events into three clusters based on their corresponding lifespans reveals a gradual deviation of the mean wave elevation and energy spectra from the New Wave theory (superposition principle) with increasing longevity of the extreme wave events.

In order to explore the role of weakly nonlinear effects in the lifetime of RWs, we compare the fully nonlinear ESBI numerical results with the weakly nonlinear CNLS simulations. Both frameworks exhibit very good qualitative agreement, with consistent changes in wave elevation patterns and directional spectral broadening, particularly observed for increasing extreme event lifespans. Interestingly, Cluster 3, which regroups the long lifespan extreme wave envelopes, reveals unique discrepancies in the RW coherence when modeled by ESBI and the CNLS. This is likely due to the limitations of the CNLS approach \citep{liu2022statistical} and of the Hilbert transformation for such wave systems. While predicting such extreme localizations analytically is to date not possible, the spatio-temporal localized and directional RW solutions \citep{qiu2016rogue,guo2020two} offer promising avenues for future investigation.

In conclusion, this work unveils a characteristic type of directional nonlinear and coherent RW structure in crossing seas, highlighting the role of nonlinear wave interaction in the formation of extreme events in colliding two-wave systems. The obtained RWs are characterized by short and long lifespans along the mean wave direction while the corresponding directional spectrum broadening is followed by a distinct dual bimodal pattern. We also confirm the sufficient role of CNLS to qualitatively simulate such long-lived freak waves. Further theoretical and experimental studies are necessary to fully comprehend and predict these nonlinear waves, beyond the limitations of the wave setup and the CNLS framework as adopted in this work.

\section{Acknowledgement}
Y.H. acknowledges the support from the Distinguished Postdoctoral Fellowship Scheme of the Hong Kong Polytechnic University (PolyU). Y.H. and J.W. show gratitude to the sponsorship provided by the University Grants Committee, Hong Kong (P0039692), PolyU, Hong Kong (A0048708), Research Institute for Sustainable Urban Development at PolyU, Hong Kong (P0042840), Department of Science and Technology of Guangdong Province, China (22202206050000278). A.C. acknowledges support from Kyoto University’s Hakubi Center for Advanced Research.

\bibliographystyle{jfm}
\bibliography{main}
\end{document}